\documentclass[twocolumn,twoside,slac_two]{revtex4}
\usepackage{graphicx}
\usepackage{fancyhdr}
\begin{document}
\title{Round Table Discussion at the Final Session of FPCP 2008:\\
The Future of Flavor Physics and CP}

%

\author{Jeffrey A.Appel}
\affiliation{Fermilab, Batavia, IL 60510 USA}
\author{Jen-Feng Hsu}
\author{Hsiang-nan Li}
\affiliation{Institute of Physics, Academia Sinica, Taipei, Taiwan}

\begin{abstract}

The final session of FPCP 2008 consisted of a round-table discussion among
panelists and audience.  The panelists included Jeffrey Appel(moderator),
Martin Beneke, George W.S. Hou, David Kirkby, Dmitri Tsybychev, Matt 
Wingate, and Taku Yamanaka.  What follows is an edited transcript of the 
session.

\end{abstract}
 
\maketitle

\thispagestyle{fancy}

 

\section{Question: What are the big questions in flavor physics at 
FPCP08?}

Jeff Appel  

Many of us from many places have to write trip reports when we get back. 
And perhaps when writing the trip reports we could start with the big 
questions in flavor physics that came up here. This is meant to help to 
you write the trip report as well as to focus the discussions to come. A 
number of topics were suggested by people who sent an email. So you can 
read them here. 

\begin{itemize}

\item{$CP$ violation in charged vs neutral $B$ decays?}
\item{Mixing induced $CP$ violation in the $B_s$ system?} 
\item{$D-\bar D$ mixing:  How soon can we measure mixing parameter 
$x$?}
\item{Spectroscopy:  What are the $XYZ$ states in the charm 
sector (counterparts in the bottom sector?)?}

\end{itemize}

I don't need to go through them one by one, but I will ask our panel 
members to begin with what among these topics they found most important; 
what they think missing from the list.  Martin why don't we begin with 
you? 

Martin Beneke

The list includes most of the hot topics discussed at this conference. 
The first two items refer to phenomena connected with $b \to s$ 
transitions, where the window to new physics is still open widest. 
However, we have learned in the past few years that the standard flavor 
theory is working quite well. The much discussed hints in the $b \to s$ 
sector are either not conclusive (second item) or possess alternative 
hadronic standard-model interpretations (first item). The actual 
observation of  $D-\bar D$ mixing is exciting as a phenomenon, but because 
of theoretical uncertainties, does not tell us much that we did not 
know before about new physics.

Matt Wingate

From the lattice QCD perspective, the most interesting thing discussed 
here was the discrepancy between the HPQCD calculation of $f_{D_s}$ and 
the experimental measurement.  The lattice result is quite sound: the 
non-strange decay constant $f_D$ is the one which requires more work, 
namely extrapolating lattice data to the physical up/down quark mass. The 
fact that $f_D$ agrees with experiment while $f_{D_s}$ does not is an 
interesting puzzle. The precision quoted for the lattice result is very 
impressive, and further details from the authors will allow other lattice 
experts to judge the quality of the fits involved.  It doesn't seem 
plausible to me that the source of the discrepancy could be blamed on the 
fourth-root hypothesis used in staggered-quark calculations.

One thing which I am investigating is: What more can be done on the 
lattice in studying $b \to s$ decays?  There are difficulties for the 
lattice here which are not present in $b \to u$ decays or neutral $B$ 
meson mixing.  Nevertheless, the $b \to s$ decays are of such great 
interest that all approaches, including lattice QCD, should be pushed as 
far as possible.  I think there are calculations we can do which will add 
to the picture.

Dmitri Tsybychev 

I just want to add that whether there is mixing induced $CP$ violation in 
the $B_s$ system will remain a hot topic for next couple of years, and 
hopefully both D0 and CDF experiments will have updates on their results; 
if not in the summer 2008, then in the fall. There is room for improvement 
on the precision of measurements of $\phi_s$ for both experiments. With 
continuing successful running of the Tevatron, both experiments plan to 
collect up to 8 $fb^{-1}$ of data. CDF already has a sample of 
3 $fb^{-1}$. Their current result is based on a data sample of only 
1.3 $fb^{-1}$. The D0 experiment has already used the full sample of 
2.8 $fb^{-1}$ available to date. 
Therefore it will be able to increase its sample only when new data are 
collected. However, D0 plans to improve the selection of $B_s$ mesons 
decaying into $J/\psi \phi$. As was already mentioned, D0 can increase the 
statistical significance of its sample by 20\% through a better selection. 
This will directly translate to an improvement of the measurement.

Additionally, a question still remains involving SU(3) or U-symmetry. D0 
constrains the strong phases involved in the $B_s$ angular analysis to the 
similar phases that appear in $B_d \rightarrow J/\psi K^*$ decays, and are 
measured at 
B-factories. The constraint is rather weak, and allows for SU(3) symmetry 
breaking, which may be as big as 10\%. Polarization amplitudes, measured 
in $B_s$ and $B_d$ decays, are compatible within measured uncertainties. 
This may 
indicate that such symmetry exists. The result on $phi_s$ does not change 
significantly if the phase constraints are removed. However there is no 
consensus whether such a constraint should be applied, and one can benefit 
from a stronger theoretical motivation.

Jeff Appel

You think that the systematic errors are not coming soon for how well you 
can do on this?

Dmitri Tsybychev 

The fit result for the case of free strong phases is provided in the D0 
article in PRL, and agrees very well within statistical uncertainty with 
result of the constrained fit.

Jeff Appel

Anybody else?  Does anybody in the audience want to add to this list? 

David Kirkby 

I think the main question in flavor physics is where the new physics is 
going to show up, if anywhere. We should remember also that there are 
certainly topics in flavor physics that have intrinsic interest: 
spectroscopy, for example. But how likely is it for new physics to show up 
there? To get the audience more involved, how about a show of hands?  
Where do you think that the new physics is likely to come from? Raise your 
hand once at which one of these four you think is the most promising. So, 
how about the first one?

Jeff Appel

You have to leave your hands long enough for count. 1, 2, 3, 4..

David Kirkby 

So how about the second one, the $B_s$? Which one of the four is the new 
physics most likely to show up? 

Rahul Sinha

They are connected. If you find $\Delta_S$ not equal to zero in $B$ mixing 
you are likely to find other signals of new physics such as a deviation in 
the small $B_s$ mixing phase among other things. They are connected, since 
$\Delta_S$ can be written in terms of the small $B_s-\bar B_s$ mixing 
phase.

David Kirkby 

The second one. What is generated from the $B_s$ system? Can we find new 
physics there? How about $D-\bar D$ mixing? Well you don't know, but 
what's 
your intuition? What's your gut feeling?

Choong Sun Kim

I thought of the story of the $D-\bar D$ mixing. There is no standard 
model 
prediction. How can you find new physics?

David Kirkby

How likely do you think you are going to find something there?

Rahul Sinha

Yes, you can measure the $D-\bar D$ mixing phase with a precision of 
about 1degree at Super-$B$, but we need 50 inverse attobarns.  

David Kirkby

How about spectroscopy? Beyond the standard model? QCD is not new physics. 

Unidentified voice

It's kind of obvious that new physics will show up there, and new 
particles can contribute to the amplitudes like penguin decays or 
$B \to \tau$ decays. In my opinion, this will be the best place to look 
for new physics.

Jose Ocariz

I agree with the previous comment that it's a necessary condition; but 
it's not sufficient. For example, if we think of item 1, I have a feeling 
that this is more or less motivated by the measurement of the different 
$CP$ asymmetry in $B \to K\pi$ decay. This is a non-controversial 
measurement, but the interpretation is not uncontroversial. There is no 
way of falsifying the standard model by this kind of measurement despite 
the fact there is potential sensitivity to contributions from 
non-standard physics.

[Comment by Tom Browder added in preparing this report:  The discussion
seemed to imply that there is no possible future resolution of this issue.
However, the isospin sum rule proposed by Gronau and Rosner is a 
model-independent test for new physics. It requires much more data ($>$ a 
factor of ten) and much more precise measurements of 
$A_{CP}(B \rightarrow K^0 \pi^0)$.]

Gerald Eigen 

Martin, you brought up the $b \to s$ transitions. I agree with you that
these are important. Since point 1 is rather general, wouldn't you rather
split them into subtopics that are associated with different points than
including them all under point 1?

Martin Beneke

I was thinking of mixing-induced $CP$ violation. Would you like to include 
$b\to s \ell\ell$?

Gerald Eigen 

Yes, and also the $b$ transitions involving $s \bar s$, like $\phi K_s$,
$\eta' K_s$, etc. The leptonic penguins clearly belong under point 1,
while the gluonic penguins fit better under point 2.

(George) Wei-Shu Hou

I know I am viewed as a fanatic, saying that fourth generation this and 
that., fourth generation for everything. I actually quite agree with what 
Martin and Jose said, and that this kind of discussion can be endless, and 
we are not going to go very far. But the converse is not true; that if you 
show that in some new physics model you can generate an observed effect, 
it would still be of interest. So [going into a short presentation] this 
result here is published in 2005, and I give you the diagram. I have said 
many times during the conference, that having a t' would bring in large 
Yukawa couplings and new $CP$ violating CKM elements. Our study was 
re-done at next to leading order in PQCD, and the effect on DCPV 
difference was not diluted. And there is another thing, that it does push 
down $\Delta_S$.
 
It's not sufficient to generate the central value of the experiment, but 
to me that is very interesting. The two things mentioned in this 
conference are of note to me. One thing.Maybe I pull this slide [from 
Derek Strom's talk] back. If you look at this, here is the Standard Model 
expectation for $\phi_s$, and here are all four different, related 
measurements. All the measured values fall to the left. And here is the 
actual published prediction from 4th generation (which is smack in the 
middle of the experiments). I already stated something like this, large 
$\sin(2 \Phi_{B_s})$, in 2005. This is on the record. At the moment, I am 
not 
a UT-fitter fan, and nobody here is. I am not on the IAC. I would have voted 
for them to be here, just for the debate. At the moment, you know, 
experimentally one can not yet say too much. It's not inconsistent with 
the Standard Model. I do point out that these numbers normally would be 
scattered (if the SM is correct), but they are not. The error bars will 
get reduced, say in next two years, from 1.35 inverse femtobarns of data, 
to 3 to 5 to 8. In the last year or two, I used to say that if the central 
value stays, I would then be willing to bet a good bottle of red wine that 
the 4th generation is real. Starting a year ago at FPCP in Slovenia, the 
data seem to be heading in this direction. Now here [another slide on 
$A_{FB}$ from Eigen's talk] is one thing that Gerald brought up but didn't 
really go through. The green line in the Belle plot is marked "C9, C10 
sign-flipped", which is equivalent to C7 sign flip. The blue line in the 
BaBar plot is for the Standard Model, almost zero, but slightly negative. 
Now the upper figure was actually shown by Dmitri [Tsybychev] in his talk. 
The blue dashed curve, is the fourth generation differential $A_{FB}$, and 
the marked red line gives roughly the lower $q^2$ bin here. So you can 
understand why the Standard Model is slightly negative and close to zero; 
because below the zero is negative. Sorry that the sign convention is 
opposite to the B-factory experiment. And above the zero is positive but 
there is a bit more negative than positive so that you get the blue zero, 
or close to zero, of the SM in the BaBar plot. But in our fourth 
generation analysis the line moves down. So the zero moves further down, 
and there is not much negative part but large positive part; so it's more 
consistent with Belle/BaBar results. And I think it was Uli who raised 
this issue, you know, complaining what is still called by experimentalists 
the C7 sign flip. This is basically a way that experimentalists say that 
there is a deviation. And this is why I stressed that I want to treat 
these things more generally, to allow complex Wilson coefficients. This 
gives the shaded area. I don't want to go into any further details. Let me 
change tone and say --- I am willing to bet a good bottle of Champagne 
now, if you want to take up the order. Why? Now this [yet another slide] 
is the standard folklore that Standard Model $CP$ violation is $10^{-20}$. 
Here is the Jarlskog invariant, and A here, the invariant CPV area, is 
like $10^-5$. But the real suppression is coming from these small masses. 
So if you put in numbers, when you normalize properly with, say, the 
electroweak phase transition temperature, you get this $10^{-20}$. Now you 
see the fourth generation does miraculous stuff here because it naturally 
has large Yukawa couplings. So if you shift by one generation, this 
$m_c^2-\mu^2$ becomes $m_t^2-m_c^2$, etc. This gives rise to a very large 
enhancement. Well, it is still a suppression factor, but the $m_b^2-m_s^2$ 
alone is the only suppression. So this gives a $10^{15}$ gain, where about 
a factor of 30 is from the $b\rightarrow s$ $CP$ violating analysis. OK, 
but the factor of 30 compared to $10^{15}$ is nothing, so long that this 
factor of 30 is not $10^{10}$, or something. So we have a very large 
enhancement factor compared to the Standard Model three generation 
Jarlskog invariant. I think this is another proof that Nature is more 
ingenious than anyone of us here. But for me, to be able to jump back to 
put the $CP$ violation within Yukawa sector to be relevant for 
baryogenesis, that's why I say I am willing to bet a good bottle of 
Champagne now, ... but only for ten people, OK? 

Choong Sun Kim

I do not know all the details of fourth generation, but I have some simple 
questions. First, as you know, and as everyone knows, this fourth 
generation neutrino mass is quite heavy, $>$ 45 GeV.  So why do we have 
such a heavy neutrino, much different from the first three generations? 
That's very strange to me. This kind of thing comes out more naturally if 
we have something like a string-inspired E(6) model, which predicts 
rather heavy vector-like quarks, unlike the fourth generation.

(George) Wei-Shu Hou

Well, there is a very simple answer to that. Vector like quarks will not 
have this enhancement. These are not masses, these are Yukawa couplings. 
Dirac masses go into the denominators, propagators and decoupling. So we 
can not have enhancement.

Choong Sun Kim

Something like Kaluza-Klein or some other excited states. I think probably 
a similar result will come out generally without a so-heavy neutrino 
problem.

(George) Wei-Shu Hou

Yea, OK. I can not argue with Kaluza-Klein. They are all legitimate, but 
this one (4th generation) is within Standard Model dynamics! Now for 
neutrinos, we firmly know there are only three light ones. But since 1998, 
as compared to 1989, we also learned that neutrinos have mass. So it's a 
much richer sector than we knew of. Furthermore, you didn't mention 
electroweak precision tests, right? There is a recent paper by Kribs et al 
(Plehn, Spannowsky and Tait), which refutes the very stringent application 
of precision test against the fourth generation in the PDG. list. So it's 
not ruled out. But whatever you say, I am just saying this more than 
ten-order-of-magnitude gain is so enormous. I use this to argue that, 
despite electroweak precision tests, even the neutrino stuff, the 4th 
generation is fairly legitimate. The thing is, when you have high scale 
$CP$ violation for baryogenesis, such as leptogenesis, you tend not to 
have a laboratory test. It's a matter of physics in the usual sense.

Taku Yamanaka 

I didn't vote for any of the four items up there. Since I am an 
experimentalist, and since I work on kaons, I will vote for kaon physics 
experiments.  The sensitivity of  a $K_L \rightarrow \pi^0 \nu 
\bar \nu$ experiment will first go down by three orders-of-magnitude, 
from O(1E-8) to O(1E-11).  Even beyond the Grossman-Nir limit, there is a 
two-orders-of-magnitude parameter space for new physics to appear.  So, do 
you want to vote for a 10 percent effect, or do you want to vote for a 
large parameter space with two orders-of-magnitude? I would vote for a 
two-orders-of-magnitude effect.

\section{Question: What are the big flavor-physics questions to come?}

Jeff Appel

There is another way to continue this discussion which is the second 
question. That is, what are likely to be the big flavor-physics questions 
after the first Tevatron or LHC signal beyond the standard model? And a 
corollary question is what would be the flavor-physics questions if we 
don't see a new signal at LHC?

The answer given for the first part is that the interesting flavor-physics 
question will depend on what you see.  However, almost anything you see 
will have multiple possible answers, multiple models which can explain it. 
This may mean that there are sensitivities to flavor physics across the 
board. In fact, I don't think a signal in a particular channel will lead 
to only one flavor-physics parameter that you want to look at. That's how 
I guess I would put it.

Tom Browder

If a signal really shows up early at the LHC, I think the big question 
will be how any new particles at LHC do not produce flavor changing 
neutral currents. The theorists will have to find brilliant ways for 
cancellations to not produce flavor changing neutral currents, not just 
produce a new model.

Jeff Appel 

So you don't think there will be big signals from LHC? I didn't mean to 
put too many words into your mouth. Anybody on the panel want to respond 
to this more ambiguous question?

David Kirkby

I think it is easy to imagine new physics at LHC where you wouldn't really 
know what to do at the Super-B-factory. So maybe the challenge to the 
audience is "Can you think of something we may find at the LHC where it 
would be unclear what to do in flavor physics?" Are there other scenarios? 
Let's talk about that.

Rahul Sinha

If you see a signal of something at the LHC, you want to make sure that 
the theoretical parameters corresponding to your favorite model/scenario, 
and that are consistent with the signal, are not actually ruled out by 
precision tests; and $B$ physics would provide a precision constraint, 
through loop contributions. Therefore, you want to make sure that $B$ data 
is consistent with the scenario and the observed signal. That is one way 
again of using flavor physics.

David Kirkby

There are strong constraints from the data we already have.

Rahul Sinha

This is not enough. As to whether the current flavor constraints are good 
enough - let me say we need to improve; we need as much improvement as 
possible. With the LHC alone, we may see a signal of new physics, but we 
may not be able to figure out what kink of new physics it corresponds to. 
Here is where flavor physics comes in, ruling out or finding consistency 
among different models given a particular signal. The better the 
precision, the better the constraints. One requires flavor physics to 
enable pinning down what is the new physics.

Martin Beneke

We discuss flavor physics in the context of the TeV scale. In doing that, 
we almost always implicitly assume that electroweak symmetry breaking is 
caused by some weak-coupling phenomena. That's not guaranteed. An entirely 
different way of seeing things would be needed if it turns out that 
electroweak symmetry breaking happens through some QCD-like 
strong-coupling mechanism. Then the flavor-physics puzzle is more severe, 
because if there is no weak coupling at the TeV scale, we would know that 
flavor physics is probing much higher scales which are disconnected from 
TeV scale. So, indirectly, one of the big flavor-physics questions to come 
and be answered is what causes electroweak symmetry breaking.

Keh-Fei Liu

I wonder if one of you could comment on neutron electric dipole moment in 
terms of its discovery potential, and if there can be some effect found in 
the next couple of years. Will the new physics be orthogonal or 
complimentary to this flavor physics?

Jeff Appel

The coupling to the neutron electric dipole moment for any of these 
questions. George?

(George) Wei-Shu Hou

You mean something specific. I actually asked Junji Hisano, the expert. 
Basically 4th generation effects can enter through loops. I think that's 
what you are referring to. So that should be studied, yes.

Keh-Fei Liu

In the program of looking for new physics, I want to see whether there is 
a discovery in a channel that would be complementary to the study here in 
flavor physics. Or, is there some orthogonal result?  

Dmitri Tsybychev 

I think it's a general problem. If you see something at the LHC, how do 
you reconstruct the underlying physics? Say that 200 models can give you 
the same signal. It will take more than just one significant deviation in 
one channel to really understand the nature of the new physics.

Rahul Sinha

Typically, one talks about the missing $E_T$ signal at the LHC. I would 
like to get an opinion as to whether flavor physics can help in pinning 
down the nature of the new physics.  With flavor-physics constraints 
included, it would be interesting to come up with signals that can help to 
say whether it is SUSY or not SUSY. 

Dmitri Tsybychev

If you have missing $E_T$, it could be anything. It could be 
supersymmetry. It could be a leptoquark. It could be extra dimensions. 
There are a number of scenarios that will result in large missing $E_T$.

Rahul Sinha

Sure. But, what is it that should be really watched out for, say, for SUSY 
or other new physics, and what kind of measurements in flavor physics can 
actually help distinguish between the kinds of scenarios. Is it possible 
to do that? Anybody?

Jeff Appel

I think the point is that too many things have missing-energy signals to 
say that this or that is the specific answer.

Tom Browder 

There is a sort of a worldwide effort, at CERN and other places. People 
are writing very thick yellow books about the connection between flavor 
physics and the physics at LHC. They do consider lots of different 
scenarios in the possible impact of all the observables in B physics. You 
may find reading these articles boring now because we don't have a new 
physics signal at the LHC to look at. But there have been pretty 
substantial efforts and a lot of papers on this.

Enrico Lunghi

I have a general comment on the first two questions. ATLAS and CMS are 
mostly "flavor-diagonal" experiments. On the one hand, they will tell us 
the mass scales and the tree-level structure of whatever new physics model 
is realized in nature. On the other hand, the quantum structure of the 
theory (e.g. loop effects) will be hardly accessible. The latter task is 
perfectly suited for flavor-physics experiments, that will act as a 
tie-breaker among the several equivalent new physics models that will 
emerge from the first analyses of LHC data. Of course, these kinds of 
studies require inputs from ATLAS and CMS. Once a few masses and processes 
are known, one can construct complete models and predict which flavor 
observables are expected to deviate from the SM predictions. It is also 
possible that ATLAS and CMS will not find any new physics. In this case, 
flavor physics (including lepton flavor violation) will allow us to access 
to much higher scales (e.g. hundreds of TeV). There are two scenarios. If 
ATLAS and CMS find TeV-scale new physics, flavor physics will help to find 
out the detailed structure of the theory. If, on the other hand, new 
physics turns out to be beyond the reach of direct production at the LHC, 
we can still explore it via super-rare processes (e.g. lepton flavor 
violation).

Choong Sun Kim

I have some unrelated questions for Hsiang-nan and Martin about the 
previous discussions. Everyone knows that we, within the standard model, 
can not calculate the $B$ to $\pi^0\pi^0$ branching fractions. Is that new 
physics? 

Martin Beneke

No.

Choong Sun Kim

Because the error is quite small. The experiment error is small.

Martin Beneke

But the theoretical error is not so small.

Choong Sun Kim

But you can explain all others except for $\pi^0 \pi^0$. Even $B$ to 
$\rho^0 \rho^0$, which has exactly same quark diagrams as $\pi^0 \pi^0$, 
can be predicted rather well. When the measurements began, it was quite 
different - theory predicted only 1/3 of the experimentally measured 
branching fraction. So I think today's value is kind of a  post-diction. 
You just changed the input parameters. Therefore, even though we think it 
is rather trivial, like the color-suppressed tree, it can be something 
else - like beyond the standard model.

Martin Beneke

We have learned that the dynamics behind the color-suppressed tree 
amplitude is very different from the naive factorization picture, and 
also understand why the theoretical uncertainties are large for this 
amplitude.

Rahul Sinha

I just want to ask something since you raised the question about 
factorization and naive factorization. Naive factorization works so well 
in $D$ decays. We all remember the classic paper of Bauer, Stech and 
Wirbel. Factorization, however, does not work so well in $B$ decays as is 
evident from data. Is there a good explanation for that?  Why does 
factorization work better for $D$ decays and not that well for $B$ decays?

Martin Beneke

I wouldn't say that this is true. In $B$ decays, we discuss many more 
challenging observables than just branching fractions of tree-dominated 
decays; such as penguin-dominated decays, $CP$ asymmetries, and strong 
phases.

Rahul Sinha

Let us just go back to branching ratios for modes like $K \pi$, $\pi\pi$ 
and ...  These things work so well in $D$ decays, but not that well in 
$B$ decays. 

Hsiang-nan Li

But I think this question does not belong to this category. I think it's 
still too early to have any concrete conclusion because currently the 
theoretical precision is just up to next-to-leading order, right? So there 
is next-to-next-to-leading order, next-to-next-to-next to leading order. 
There is a long way to go.

\section{Question: What are the connections between observations in the 
quark and lepton sectors?}

Jeff Appel

This is pretty technical for the round-table level of discussion. I guess 
I'd like to move on to our next question. I don't have a lot of questions. 
Don't get too scared. I wonder about the connection between the flavor 
observations in the quark and lepton sectors. Do we understand these? Or, 
do we have to wait to get to Plank scale to figure it out.

Choong Sun Kim

The $\sin(\theta_{12})$ in neutrino-sector mixing and $\sin(\theta_{12})$ 
in the quark 
sector, now adding up those 2 mixing angles comes up to about 45 degrees. 
It could be an accident. Or maybe there is some kind of connection between 
the quark sector and the lepton sector. People say it's complementarity, 
something like that. Quark-lepton complementarity. Maybe there is some 
reason behind it, or is it an accident?

\section{Question: Is there a flavor-physics community, and if so, has 
it articulated its case well enough?}

Jeff Appel

One reason why I put this question in here is to address the nature of 
this conference and our community. I use the singular form, our community, 
the flavor-physics community which covers quarks and leptons. This is the 
physics we have discussed at FPCP 2008. Have quarks and leptons been 
brought together at this meeting more strongly than in the past because of 
$CP$ violation only, or there is something more fundamental that makes 
them part of the same community? And if so, has this community articulated 
the case for support of both axes strongly enough? I am thinking of the 
priorities that have been expressed in the United Kingdom and in the 
United States. We also have heard about the delay in kaon physics at 
J-PARC, and so on.

Taku Yamanaka 

Well, let me first speak about the situation in Japan. The High Energy 
Physics Committee in Japan, of which I am also a member, wrote up a report 
on what to do in the future. In that report, we stated two things. One is, 
approach the high energy frontier, including LHC and ILC etc. We also 
stated that the intensity frontier, especially flavor physics, is 
important. This is especially true because in Japan we have Belle and the 
neutrino program. The experiments are very popular and are being 
supported. J-PARC is the key facility for neutrino and kaon experiments. 
Even the people pushing for the ILC are supporting the J-PARC program, 
because if J-PARC fails, then there is no linear collider. From the 
viewpoint of the funding agency, that's very clear. 

If the question is, is there a flavor-physics community in Japan, the 
answer is yes. The people working on kaons, $B$ physics, and neutrino 
physics, experimentalists and theorists, have joined forces and won a 
"Grant-in-Aid for Scientific Research on Priority Areas", titled "New 
Developments of Flavor Physics". The project is supported for 6 years, and 
the fund is being used for building the T2K and Opera experiments, the 
J-PARC kaon experiment, $B$ physics at CDF, and Belle. We get together 
every year to have a small workshop to present all the new findings. This 
is really making a close community of people ranging from young students 
to older professors working on various experiments and theories, all on 
flavor physics.

Martin Beneke

It may be unpopular to say this, but talking to people outside and even 
within the flavour physics community, one may get an impression that 
flavor physicists had their chance to find new physics. They did not, 
so it is time to move on to the next thing - LHC physics. If something 
shows up there, then we can go back to flavor physics to try to sort 
things out. We may be blamed ourselves for that because we have been 
talking too much about new physics and obscure 2$\sigma$ effects, and 
didn't succeed to create interest in the intrinsic physics itself, in the 
phenomena. 

I am fascinated and mystified how neutrino physics is succeeding in 
this respect -- measuring a mass matrix in the lepton sector, which 
is after all not so different from measuring the CKM matrix. And there 
is even less prospect of discovering new physics by determining 
$\theta_{13}$ than there is in $V_{ub}$!

Jeff Appel

There is an interesting corollary to the way you put it. In terms of 
selling the physics these days, one tries to sell physics as 
"paradigm-changing" discoveries. What is the argument you would make to 
sell our physics, whatever it is? The first thing one looks for is what 
people call paradigm-changing discoveries, right? How would you sell the 
physics that you are talking about in the world?

Martin Beneke

Once neutrinos have masses, there is no paradigm change in measuring 
mixing angles or even CP violation. Nevertheless, there is some 
intrinsic interest in investigating neutrino properties, because 
neutrinos are considered mysterious, while quarks are not. In any case, 
returning to your question, advertising paradigm change is dangerous, 
since paradigms usually change by observations that come unexpectedly, not 
because of a systematic search.

Jeff Appel

Martin, that's in fact exactly right. The neutrinos are interesting 
because we were surprised. And it's more a matter of surprise which sells 
newspapers, rather than continuing to observe the things we expected to 
observe, the so called standard model. That doesn't sell newspapers. The 
articles we read about 600 physicists "failing" to find this, "failing" to 
find that. We have a problem selling the more precision measurement and 
standard things.

Bruce Yabsley

You asked how do you sell something more subtle when someone else is 
pushing the paradigm-changing. The answer is: it's damn hard! If you send 
your children to their grandparents, and the grandparents feed them candy, 
they are attacking kids' weakness. I am sorry. We always push that this 
would change the world, whatever. Think of all that has been happening in 
spectroscopy, some of the most interesting stuff to come up in the $B$ 
factory. I don't believe that it's nuclear physics. 

Jeff Appel

And it's interesting because it was a surprise? Or interesting for another 
reason?

Bruce Yabsley

Again, it's interesting because it's a surprise. Now, if we get into a 
position where we discover something that is both surprising and 
interesting!  Maybe we have to spend a few years in training on how to
talk to guys from the newspapers. Maybe we just do.

David Kirkby

Maybe one way to answer your question is to look at the nuclear physics 
community because they are, at least in the US, well funded; and what they 
are doing is not so different from spectroscopy in heavy quarks. 

Rahul Sinha

The fact is that we initially set up the $B$ factories to test the CKM 
hypothesis. We have succeeded; we have done that. We have not only 
succeeded in doing that, but we have learned a lot more. We have new 
resonances and many puzzles about them. This is at the very least 
"surprising". So in that sense, there is no way to say that we have not 
actually had very good physics output. Somehow, $B$ physics efforts have 
become the victim of various constraints dictating the directions in 
physics, e.g. our desire to find a way probe the Plank scale as fast as 
possible.

Eli Rosenberg  

Let me say something that has already been said. The first slide you put 
up there. It all had to do with where new physics is going to be found. 
You already brought in the concept that to sell anything, it has to be 
something new. And on your second slide, the reaction to what happens in 
the Tevatron and LHC. God help this field if nothing is found in those 
places. This conversation becomes entirely irrelevant simply because we 
have oversold the idea that we have to find something new. Now I have a 
feeling that if we went back to 30, 40 or 50 years ago, when particle 
physics was a virgin, people were working on precision measurements of 
electromagnetic interactions. We must have felt exactly the same way you 
are feeling in this room now - that somehow we were undervalued by looking 
at things where you could make precise measurements. And the real argument 
is, we are working in the area where you can make precise measurements, 
where you can look for new things like lepton flavor violation. We'd like 
to measure $D$ mixing because we didn't expect to see much of it. It's 
interesting, and it has intrinsic interest of its own, period. Whether 
it's going to be something new or not, that is a different issue. Now, how 
you sell that to our funding agencies is where the problem seems to come 
in. The same thing happens in the $K$ meson sector. The $K$ meson sector 
had a resurgence at one point after being pushed down for a long time. So 
this has been a continuing problem. But I think part of the problem is 
that we have gotten so big and so expensive that we oversell everything. 
The field as a whole has oversold everything. This is what you have to do. 
That's why you read the headline about 600 physicists failing to find this 
or that; because we said we were going to find it. You know, we sold the 
SSC as if we could do everything except cure baldness. So I think we just 
have a PR-reality problem about what science is about.

(George) Wei-Shu Hou

I would like to make several remarks touching on all that has been said. I 
think that on neutrino physics, I held back on one question that I used to 
ask. If I take $V_{ub}$, it's very hard to extract, correct? But if we 
take the $V_{ub}$ analogy, because neutrino people have had ten 
spectacular years, this is in part because of the very large mixing 
angles. They could not ordain that, right? So if I take $V_{ub}$ or even
$V_{cb}$, our $\theta_{13}$ or $\theta_{23}$ for neutrinos, I don't see a 
program yet to measure something of that strength for $\theta_{13}$. They 
are entering a hard time. Without that (a large $\theta_{13})$, forget 
about $CP$ violation in the lepton sector. OK, Majorana neutrinos, 
(neutrinoless) double beta decay, there is always some discovery 
potential. But they are not really doing better than we are. I don't 
know how, in the last ten minutes, we entered such a very gloomy mood. I 
think we actually have a good situation. The LHC is starting. The Tevatron 
is still working very hard. We are seeing things here and there. We are, 
of course, used to seeing things disappear from before our eyes. But 
that's how it is; right? Seeing something emerge and then disappear; 
hoping that one of them is true. And I think this mixing-dependent $CP$ 
violation phase (in $B_s$) is of course the way to go. But, maybe we are 
overselling it. It may be a PR problem, but we do have genuine 
indications, not just challenges. So I like what Enrico said, I think at 
LHC, ATLAS and CMS analyses will find the scale. The New Particles will 
likely be extrinsic to flavor. However there is also LHCb, right? I guess 
we go back to Question 2. What if we see nothing beyond the Standard 
Model? Maybe we see the Higgs, maybe we don't. But if we see nothing, and 
LHCb will measure $\sin(2 \phi_s)$ to plus or minus 4 percent, no charged 
Higgs, no SUSY. Then no matter what PR we do, you can not get the next 
big machine.

But I am optimistic about both LHC proper, the high energy frontier, and 
LHCb also. I am also a full supporter of the Super KEKB or Super 
$B$-factory. Because it's really a PR question again. A Super $B$-factory 
is a multi-purpose facility. And speaking from Asia, I am even more 
supportive of this. Asia is rising. It has the population, etc. I would 
fully support it even just only on that account. That it is a project to 
work on, to go forward. And if not a discovery at this stage, then there 
will be a discovery at the next stage. To me, Super KEKB is a regional 
cooperation concern.

Jeff Appel

In order to move to a more positive direction, perhaps there are other 
questions people would like to address to the panel, or to each other 
before I get to my last question?

Bruce Yabsley

Just inspired by the previous discussion, I would note that when the LHC 
turns on, the field is going to undergo a kind of basic change. The kind 
of information we are using to decide what studies to do, and how to do 
it, is going to change. Now, the moment someone puts a preprint on hep-ph, 
everyone at the $B$-factories, drops everything they are doing to pursue 
the suggested analysis, or something like this. What's going to happen now 
is that we are going to get some sign of a particular mass scale. And 
that's going to have an influence on the things we should be studying. But 
the influence is going to be a hundred percent, because it's going to 
determine what the new physics is, presumably, and more than one model 
will be possible. Here my question is: do we have a mechanism so that we 
have a while to think about how we are going to be influenced by the new 
information. Or, are we going to be driven by some prejudice that what we 
see, is what we know, and so suddenly everyone rushes in particular 
directions like ten-year-olds playing soccer. The situation really is 
going to change. I am wondering if we thought forward to what happens when 
we see data. 

Jeff Appel

I think there is plenty of evidence that the community is a very good at 
rushing together in singular directions, in effect, ten-year-olds playing 
soccer as you called it. We also did this when the $J/\psi$ was 
discovered. Every experiment asked if they could see evidence of that 
signal.

Enrico Lunghi

I would like to make a comment on the impossibility of a null result at 
the LHC. In fact, unitarity tells us that either we'll find at least one 
Higgs particle. Otherwise, some other phenomena have to happen (e.g. 
strongly-interacting vector bosons, new strong resonances, ...). In any 
case, even if just a SM-like Higgs is found, we still need a linear 
collider to study its properties. I don't think that we can really go 
there and see nothing.

Eli Rosenberg 

Because you are so convinced, and we have been told that we will. And if 
we see nothing, it is the most fascinating physics result of all - except 
that it will kill the field. Aside from that, ... I rest my case. Then why 
do we have to build it? Because we are going to see something.

Choong Sun Kim

I have one question not related to politics or anything like that. We know 
that George proposes a fourth generation. But if there is a fourth 
generation, it is supposed to violate unitarity in three generations 
because, effectively, the 3 by 3 part of the larger 4 by 4 matrix is 
non-unitary. OK. What I want to ask is another thing, about gamma in the 
unitary triangle. The gamma or alpha measurement is not actually measuring 
gamma or alpha. It's beta plus gamma, for example. So my question is, is 
it possible that LHCb, or Super-B, or any future B-factory can find the 3 
by 3 CKM matrix non-unitary? Is it possible to find non-unitarity or not?

Jeff Appel

Yes, anybody working in $B$ factories would say yes, you can find the 
triangle does not close. It is not unitary and there is going to be 
something else, some new physics.

Chung-Sun Kim

The measurement of gamma or alpha is from the $\pi-(\alpha+\beta)$. So by 
definition, you are just taking the angles as from a triangle.

Jeff Appel

The sides have to work too, right?

Eli Rosenberg 

Gamma, perhaps from a Dalitz analysis, maybe from the $\Upsilon(5S)$. Does 
that beta plus gamma match the beta plus gamma you get when they 
interfere? That's the test. And that's equivalent to test unitarity; 
whether the standard model is working. That's what is, in short hand, 
called alpha. I agree with you. Nobody measures alpha, but measures
$\pi-(\beta+\gamma)$.

Rahul Sinha

The one thing to note is that $\gamma$, or one of the angles measured, has 
to be from outside of the $B_d$, to see a breakdown of unitarity. It is 
well known that if you measure the three phases using just $B_d$, the 
effect of new physics will cancel out. In addition, the other thing one 
can do is to measure both sides and the angle and then check if the 
triangle closes.

Jeff Appel

The same triangle.

Choong Sun Kim

The measurement of $\gamma$ or $\alpha$ is from $\pi-(\alpha+\beta)$
or $\pi-(\beta+\gamma)$. So, by definition, you are just forcing a 
triangle if you do not measure the 3 angles independently. Also, beta from 
$B \rightarrow J/\psi K_s$ can effectively include new physics, too.

Rahul Sinha

Yes. But, you can also measure gamma, outside the $B_d$ system.  There is 
a method to measure it using $B_s \rightarrow DK$. If you do that, then 
there is no 
problem. You can detect the breakdown of unitarity without measuring the 
side.

Jose Ocariz 

Another way of saying it is that you are measuring 4 parameters with 10 
observables. If you have no consistency, you have no unitarity.

\section{A final question.}

Jeff Appel

We are reaching the end of our scheduled time. I do want end on one new 
question, our last question, which is "How can we thank our hosts enough 
for their hospitality, the careful and caring organization of this 
meeting, their holding back the worst of the rains for the excursion, and 
clearing the sky as well for the highest level of FPCP banquet ever held? 
So, I think we should close this session and thank our hosts very much.

(applause)

And I personally want to thank the panelists and all of you in the 
audience for the very stimulating discussion. Good luck on your trips 
home, whether near or far. And, again, thank you all.

\bigskip 




\end{document}